\documentclass[a4paper,english,aip,jcp,unsortedaddress,numerical,preprint,a4wide]{revtex4-1}
\usepackage[T1]{fontenc}
\usepackage[latin1]{inputenc}
\usepackage{booktabs}
\usepackage{amsmath}
\usepackage{graphicx}

\makeatletter

\providecommand{\tabularnewline}{\\}

\usepackage{babel}
\makeatother

\begin{document}

\title{Semi-empirical correction of\emph{ ab initio }harmonic properties\\
by scaling factors: a validated uncertainty model\\
for calibration and prediction}

\author{Pascal Pernot}

\affiliation{Laboratoire de Chimie Physique, Univ Paris-Sud, Orsay, F-91405}

\affiliation{CNRS, UMR8000, Orsay, F-91405}

\email{pascal.pernot@u-psud.fr}

\author{Fabien Cailliez}

\affiliation{Laboratoire de Chimie Physique, Univ Paris-Sud, Orsay, F-91405}

\begin{abstract}
\noindent Bayesian Model Calibration is used to revisit the problem
of scaling factor calibration for semi-empirical correction of \emph{ab
initio} harmonic properties (\emph{e.g.} vibrational frequencies and
zero-point energies). A particular attention is devoted to the evaluation
of scaling factor uncertainty, and to its effect on the accuracy of
scaled properties. We argue that in most cases of interest the standard
calibration model is not statistically valid, in the sense that \emph{it
is not able to fit a set of experimental calibration data within their uncertainty
limits}. This impairs any attempt to use the results of the standard
model for uncertainty analysis and/or uncertainty propagation. We
propose to include a stochastic term in the calibration model to account
for model inadequacy. This new model is validated in the Bayesian
Model Calibration framework. We provide explicit formulae for prediction
uncertainty in typical limit cases: large and small calibration sets
of data with negligible measurement uncertainty, and datasets with
large measurement uncertainties.
\end{abstract}
\maketitle

\section{Introduction}

One considers generally two types of uncertainty, arising either from
random errors or from systematic errors. In quantum computational
chemistry, random uncertainties, such as those arising from non-zero
convergence threshold, have been shown by Irikura \emph{et al.}\citep{Irikura04}
to be negligible. The major uncertainty sources are \emph{biases}
due to basis-set and/or theory limitations. For quantum chemistry
to be predictive, \emph{i.e.} to be able to \emph{predict observables
with confidence intervals}, these biases have to be corrected. A common
way to do this is by semi-empirical corrections, \emph{i.e.} corrections
by additive or multiplicative factors calibrated on sets of experimental
data.\citep[See e.g.][]{Pople81,Curtiss98,Curtiss00,Wilson01,Curtiss07} 

Semi-empirical corrections of \emph{ab initio} results by linear scaling
are efficient for many observables. It is often overlooked that semi-empirical
corrections are statistical operations, and as such, accompanied by
an uncertainty which has to be considered in the uncertainty budget
of model predictions, of which it is liable to be a major contribution.
A sound uncertainty budget for these corrections is important in many
circumstances. For instance, it is acknowledged that ZPE is a major
source of uncertainty in thermochemistry with chemical accuracy.\citep{Feller06,Feller07,Feller08}
A good evaluation of ZPE prediction uncertainty is therefore essential
for the assessment of the accuracy of computed thermochemical properties.
In another field, infrared spectral fingerprinting, confidence intervals
on corrected vibrational frequencies could help to ascertain the identification
of spectral features.\citep{Sinha04,Andersson05,Simon07,Bouteiller09,Poully09,Barone09,Biczysko09}
Estimation of uncertainty on computational chemistry results is also
of paramount importance for their transfer in multiscale chemical
modeling.\citep{Pancerella03,Frenklach07} As quantum computational
chemistry is at the lowest scale of chemical simulation, uncertainty
on its results has to be carefully propagated to the higher scales
in order to get quantified predictions. An example is the use of computational
thermochemistry to predict the rates of reactions that could have
a direct impact on macroscopic observables in combustion simulations.\citep{Androulakis06}

The concept of \emph{Virtual Measurement} has been introduced by Irikura
\emph{et al.},\citep{Irikura04} with the aim to recast model outputs
in the standardized uncertainty management framework established for
experimental measurements in the \emph{Guide to the Expression of
Uncertainty in Measurement} (also known as ''the GUM'').\citep{GUM}
To be a Virtual Measurement, a model output has to be qualified by
a standard uncertainty or confidence interval. 

\begin{figure}
\noindent \begin{centering}
\includegraphics[clip,width=12cm]{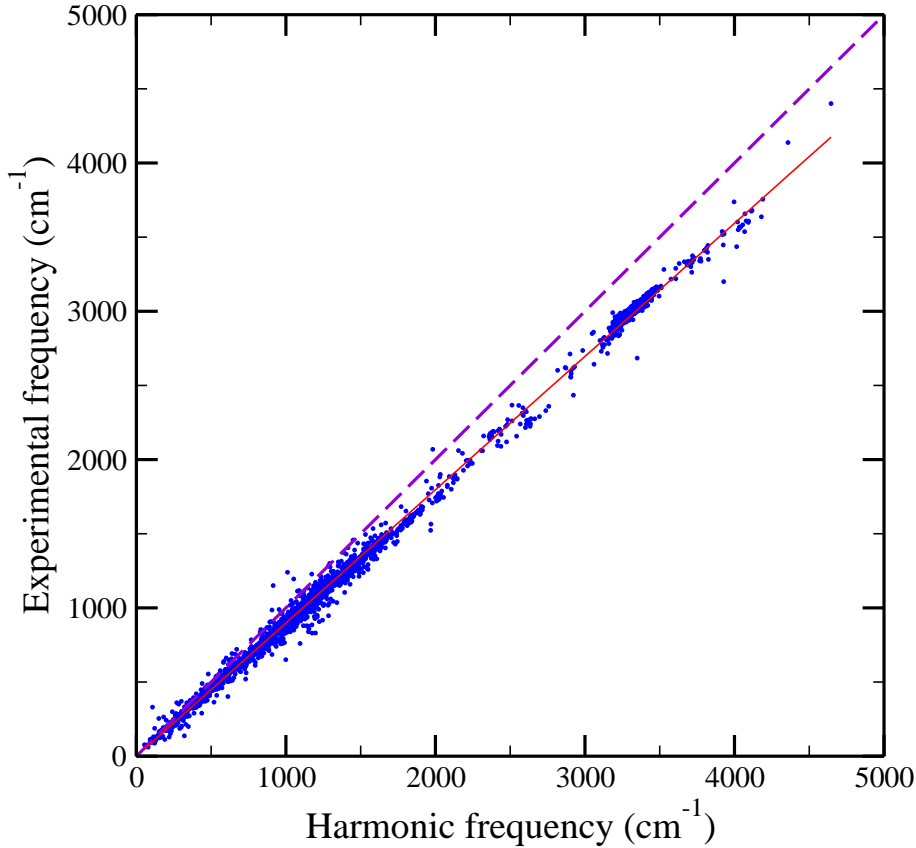}
\par\end{centering}

\caption{\label{fig:Correlation-plot}Correlation plot between calculated harmonic
frequencies $\omega_{i}$ and measured frequencies $\nu_{i}^{exp}$
for a set of vibrations extracted from the CCCBDB for the HF/6-31G{*}
combination of theory/basis-set (dots). The full line is the regression
line $\nu_{exp}=s\omega$; the dashed line is a visual aid to appreciate
the bias.}

\end{figure}

In a recent article (hereafter IJKK09), Irikura \emph{et al.}\citep{Irikura09}
address the problem of uncertainty evaluation for scaled zero-point
energies (ZPE), in the continuity of their 2005 paper (hereafter IJK05)
on vibrational frequencies.\citep{Irikura05} Scaling of harmonic
vibrational frequencies%
\footnote{In the following ''vibrational frequency'' could be replaced by
any other observable calculated at the harmonic level. %
} is an important example of semi-empirical correction method in computational
chemistry, where estimation of a vibrational frequency $\nu$ is obtained
by multiplying the corresponding harmonic vibrational frequency $\omega$,
routinely calculated by computational chemistry codes, by an empirical
scaling factor $s$ (Fig. \ref{fig:Correlation-plot})\begin{equation}
\nu=\omega\, s.\end{equation}
Optimal scaling factors have been computed for numerous sets of
theory/basis-set combinations.\citep{Scott96,Wong96,Merrick07,cccbdb} 

More sophisticated scaling schemes have been designed to increase
the precision of semi-empirical corrections. They make use of frequency-range
or mode adapted scaling factors for frequencies,\citep{Merrick07,Bouteiller09}
or internal coordinate adapted scaling factors for force constants.\citep{Blom1976,Rauhut95,Baker1998}
In all cases, the scaling factors are optimized to reproduce at best
a set of experimental data, and are affected by a calibration uncertainty,
which depends on a few factors, as the size of the calibrartion set
and the precision of the data within. We focus in the following on
the importance of this calibration uncertainty and concentrate on
the widely used uniform scaling factors (\emph{i.e.} a single scaling
factor for all frequencies), without loss of generality.

In the majority of publications about scaling factors, two summary
statistics are provided for each theory/basis-set combination: the
optimal scaling factor and the root mean squares deviation, characterizing
the average distance between experimental and corrected values estimated
on the calibration dataset. From a reference dataset of experimental
$\left\{ \nu_{exp,i}\right\} _{i=1}^{N}$ and calculated $\left\{ \omega_{i}\right\} _{1=1}^{N}$
vibrational frequencies, the optimal scaling factor obtained by the
least-squares procedure is\begin{equation}
\hat{s}=\sum\omega_{i}\nu_{exp,i}/\sum\omega_{i}^{2}\label{eq:optim_s}\end{equation}
and the quality of the correction is estimated by the root mean squares
(rms) value \begin{align}
\gamma & =\left(\frac{1}{N}\sum\left(\nu_{exp,i}-\hat{s}\omega_{i}\right)^{2}\right)^{1/2}.\label{eq:rms}\end{align}
To our knowledge, these values have not been explicitly used for uncertainty
propagation, but the rms $\gamma$ provides an estimate of the residual
uncertainty resulting from the scaling correction (''something like
the target accuracy'',\citep{Pople99} or ''a surrogate for uncertainty''
according to Irikura \emph{et al.}\citep{Irikura05}), and is used
as a criterion for theory/basis-set selection.

Acknowledging that scaling factors obtained by calibration on experimental
datasets are uncertain, Irikura \emph{et al.}\citep{Irikura05,Irikura09}
proposed that (i) this uncertainty is the major contribution to prediction
uncertainty using the scaling model; and (ii) prediction uncertainty
is proportional to the calculated harmonic property (frequency or
ZPE). These authors argue also that scaling factors are accurate to
only two significant figures, and that all other studies overstate
their precision by reporting them with four figures. This approach
has been adopted by the National Institute of Standards and Technology
(NIST) and put into practice in the Computational Chemistry Comparison
and Benchmark DataBase (CCCBDB),\citep{cccbdb} section XIII.C.2,
where scaling factors are provided with uncertainties derived according
to the procedure of IJK05/IJKK09. These results can also have
a direct impact on the criteria to define the best basis/method level
of theory for a given observable. 

In the present paper, we revisit the problem of scaling factor calibration
and properties prediction through the Bayesian Model Calibration framework,
reputed for providing consistent uncertainty evaluation and propagation.\citep{Kennedy01,Gregory05,GUMSupp1}
Section \ref{sec:Methods} presents the methodological elements used
for calibration and validation procedures, which are applied to a
few representative vibrational frequency and zero point vibrational
energy datasets and compared to the approach by IJK05/IJKK09 in Section
\ref{sec:Applications-and-discussion}. We point out a statistical
inconsistency in this approach, the main consequence being a much
too large scaling factor uncertainty, from which misleading conclusions
can be derived. A set of recommendations for reliable uncertainty
estimation of scaled properties is provided in the Conclusion. Bayesian
calculations used in this study are fairly standard and straightforward,
but for the sake of completeness and for readers unfamiliar with statistical
modeling, detailed derivations are provided in the Appendix.

\section{Methods\label{sec:Methods}}

In the following sections, we present the calibration procedure for
uniform scaling factors of vibrational frequencies, but it can be
easily transposed to any other property usually computed at the harmonic
level and corrected by a multiplicative scaling factor (ZPE, entropy...).
It is also straightforward to transpose this procedure to semi-empirical
correction schemes involving multiple frequency-adapted scaling factors.

\subsection{Scaling factor calibration}

Considering a measured frequency $\nu_{exp}$, one can assume that
it is related to the true or exact value $\nu_{true}$ by\begin{equation}
\nu_{exp}=\nu_{true}+\epsilon_{exp}\end{equation}
where $\epsilon_{exp}\sim N(0,u_{exp}^{2})$ is a normal random variable,
centered at zero with standard uncertainty $u_{exp}$, which represents
the measurement error. 

Calculated harmonic vibrational frequencies $\omega$ are also affected
by random errors, related to numerical convergence defined by non-zero
thresholds and the choice of starting point in geometry optimization,
and to non-zero thresholds in wave-function optimization.\citep{Irikura04}
It has been shown that these uncertainties are negligible when compared
to the measurement uncertainty $u_{exp}$.\citep{Irikura04} In the
following, one can thus assume that, for one choice of theory/basis-set,
the harmonic vibrational frequencies are computed without significant
uncertainty.

\subsubsection{The standard calibration model}

If one makes the hypothesis of a linear relationship between $\nu_{true}$
and $\omega$ , as popularized by Pople \emph{at al.},\citep{Pople81}
the \emph{standard calibration model} is \begin{equation}
\nu_{exp,i}=s\omega_{i}+\epsilon_{exp,i}\,,\label{eq:simple model}\end{equation}
where one considers a set of $i=1,\, N$ frequencies. For a single
frequency, there is an optimal scaling parameter $s_{i}=\nu_{exp,i}/\omega_{i}$.
As $\nu_{exp,i}$ is uncertain, with standard uncertainty $u_{exp,i}$,
the value of $s_{i}$ cannot be known exactly and has a standard uncertainty
$u_{s_{i}}=u_{exp,i}/\omega_{i}$. For a calibration dataset with
uniform measurement uncertainty $u_{exp}$, it can be shown that the
optimal value for $s$ is given by the least squares solution $\hat{s}$,
Eq. \ref{eq:optim_s}, and its standard uncertainty by $u_{s}=u_{exp}/\sqrt{\sum_{i=1}^{N}\omega_{i}^{2}}$
(\emph{cf.} Section \ref{sub:Case-of-very}).

Applicability of this formula is subject to one condition: the model
(Eq. \ref{eq:simple model}) has to be statistically valid, which
means that the residuals $\left\{ \nu_{exp,i}-\hat{s}\omega_{i}\right\} _{i=1}^{N}$
should have a normal distribution centered on zero, with variance
$u_{exp}^{2}$. Normality is not always verified,\citep{Wong96} but
most important, the variance condition is violated in most cases where
precise data are used for calibration: the linear model (Eq. \ref{eq:simple model})
is typically unable to reproduce a given set of measured frequencies
\emph{within their measurement uncertainty}. Consequently, the width
of the distribution of residuals is dominated by model misfit instead
of measurement uncertainty ($\gamma\gg u_{exp}$), which invalidates
the distributional hypothesis of the standard calibration model (Eq.
\ref{eq:simple model}). In these conditions, this model \emph{should
not} be used to infer $u_{s}$, the uncertainty of $s$. Note that
this is the key point to explain statistical inconsistencies in IJK05/IJKK09,\citep{Irikura05,Irikura09}
as will be discussed later.

\subsubsection{An improved calibration model}

An option to solve this problem would be to search for better \emph{ab
initio} methods, able to reproduce experimental properties within
their measurement uncertainties. This is an active research area which
is out of the scope of the present study \citep[See e.g. ][]{Bowman86,Wright01,Csonka05,Karton08,Barone09,Basire10}.
Considering the practical interest of correction by scaling factors,
we rather focus on restoring statistical consistency by improving
the calibration model.

\begin{figure}
\noindent \begin{centering}
\includegraphics[clip,width=12cm]{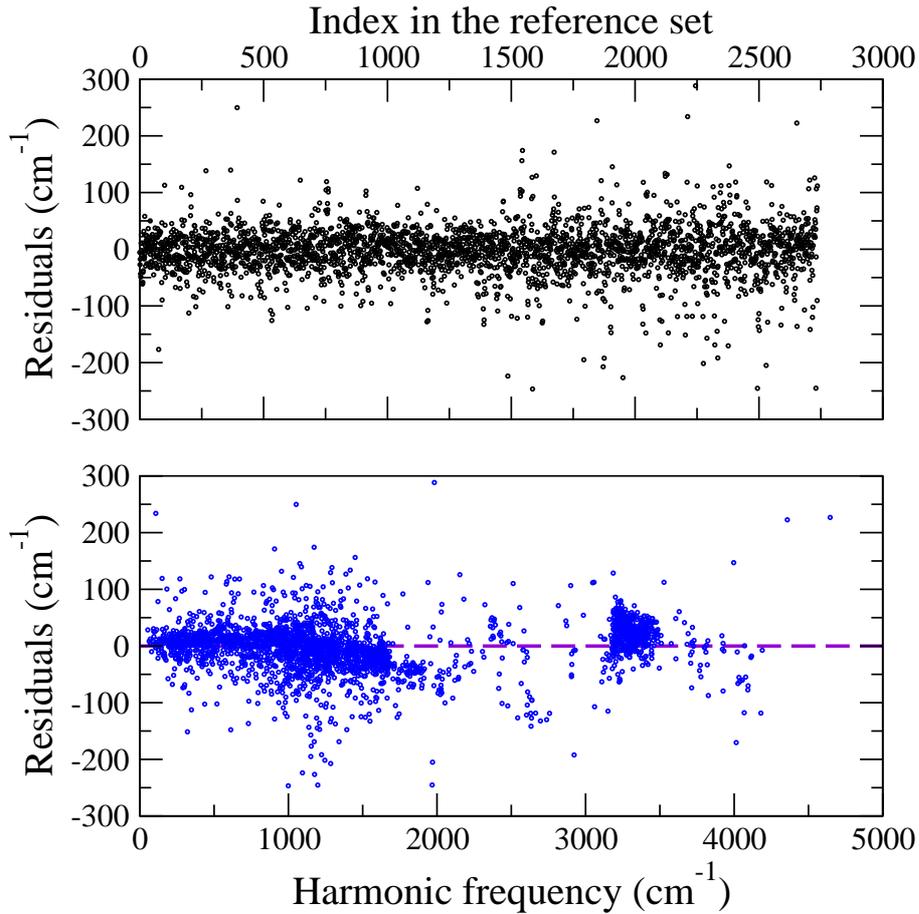}
\par\end{centering}

\caption{\label{fig:Residues}Residuals between calculated harmonic frequencies
$\omega_{i}$ and measured frequencies $\nu_{i}$ for a set of vibrations
extracted from the CCCBDB for the HF/6-31G{*} combination of theory/basis-set
(dots). Bottom: residuals as a function of $\omega$. In order to
suppress the grouping effect linked with frequencies, the residuals
were also plotted as a function of their order in the reference set
(top). }

\end{figure}

Observing the apparent randomness of the residuals $\left\{ \nu_{exp,i}-\hat{s}\omega_{i}\right\} _{i=1}^{N}$
(Fig. \ref{fig:Residues}), we consider that the model misfit is not
deterministically predictable. A solution to preserve a statistically
valid linear scaling model is to introduce an additional stochastic
variable $\epsilon_{mod}$ to represent the discrepancy between model
and observations \begin{equation}
\nu_{exp,i}=s\omega_{i}+\epsilon_{mod}+\epsilon_{exp,i}.\label{eq:calibration model}\end{equation}
This equation is similar to the basic statistical model introduced
by Kennedy and O'Hagan\citep{Kennedy01} for Bayesian Calibration
of Model Outputs. The discrepancy variable $\epsilon_{mod}$ could
formally depend on $\omega$, but we observed on representative datasets
that the residuals between modeled and observed frequencies are not
markedly frequency dependent (Fig. \ref{fig:Residues}).%
\footnote{From Fig. \ref{fig:Residues}, one could consider splitting the frequency
range in two parts (for instance below and above 3000 cm$^{-1}$).
This would enable to improve somewhat the fit (\emph{cf.} the rms
in Table \ref{tab:full_set}) and the homogeneity of the residuals,
but this would not resolve the model inadequacy issue. The Bayesian
calibration method would then have to be applied to two separate scaling
factors, instead of one.%
} Therefore $\epsilon_{mod}$ is considered null in average, with unknown
variance $u_{mod}^{2}$:\begin{equation}
\epsilon_{mod}\sim N(0,u_{mod}^{2}).\end{equation}
The new calibration model (Eq. \ref{eq:calibration model}) depends
on two unknown parameters, $s$ and $u_{mod}$.%
\footnote{Similar equations are also obtained for the calibration of stochastic
models (\emph{e.g.} based on Monte Carlo simulations) to experimental
values,\citep{Gregory05} but in these cases $u_{mod}$ is considered
to be known.%
}

\begin{figure}
\noindent \begin{centering}
\includegraphics[clip,width=12cm]{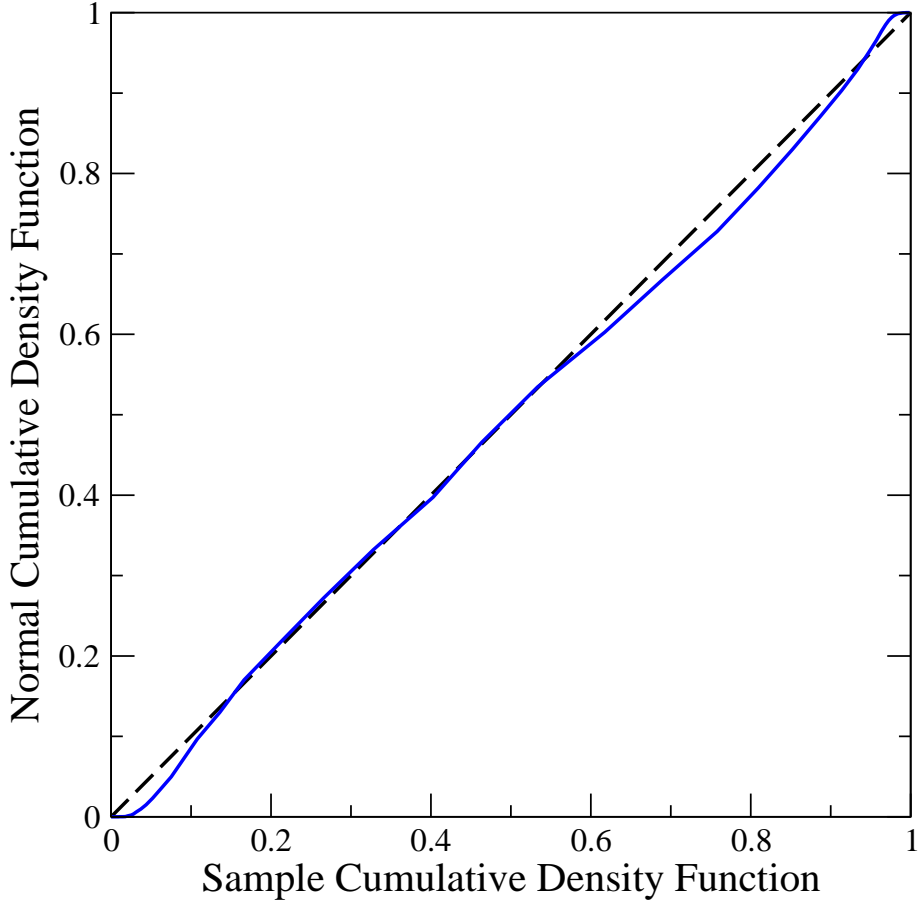}
\par\end{centering}

\caption{\label{fig:qqplot}Plot of the cumulative density function (CDF) for
the residuals (same as in Fig. \ref{fig:Residues}) against a normal
CDF shows that globally there is very little deviation from normality
in this dataset.}

\end{figure}

\subsection{Model predictions and uncertainty propagation}

The new stochastic prediction model used within the calibration model
(Eq. \ref{eq:calibration model}), \begin{equation}
\nu=s\omega+\epsilon_{mod},\label{eq:stoch_pred}\end{equation}
is linear with respect to uncertain variables $s$ and $\epsilon_{mod}$,
and one can use standard uncertainty propagation rules\citep{GUM}
to estimate the average value and variance of predicted frequencies:
\begin{align}
\overline{\nu} & =\overline{s}\,\omega\\
u_{\nu}^{2} & =\left(\frac{\partial\nu}{\partial s}\right)_{s=\overline{s}}^{2}u_{s}^{2}+\left(\frac{\partial\nu}{\partial\epsilon_{mod}}\right)_{\epsilon_{mod}=0}^{2}u_{mod}^{2}\\
 & =\omega^{2}u_{s}^{2}+u_{mod}^{2},\label{eq:UP_glob}\end{align}
where $\overline{s}$ denotes the average value of the scaling factor,
and $u_{s}^{2}$ its variance. 

In order to provide evaluated predictions of vibrational frequencies,
we need therefore to estimate $\overline{s}$, $u_{s}^{2}$ and $u_{mod}^{2}$
from a calibration dataset. This is done in the next section, using
Bayesian Model Calibration.

\subsection{Bayesian Model Calibration (BMC)}

\subsubsection{General case}

Starting from the calibration model (Eq. \ref{eq:calibration model}),
one derives the expression for the \emph{posterior} probability density
function (pdf) of the parameters, given a set of $N$ measured and
calculated frequencies (details of derivation are provided in Appendix
\ref{sub:Taking-measurement-uncertainties}) \begin{eqnarray}
p\left(s,u_{mod}|\left\{ \nu_{exp,i},u_{exp,i},\omega_{i}\right\} _{i=1}^{N}\right) & \propto & \frac{1}{u_{mod}\prod_{i=1}^{N}\sqrt{u_{mod}^{2}+u_{exp,i}^{2}}}\times\nonumber \\
 &  & \exp\left(-\sum_{i=1}^{N}\frac{\left(\nu_{exp,i}-s\omega_{i}\right)^{2}}{2\left(u_{mod}^{2}+u_{exp,i}^{2}\right)}\right).\label{eq:full_pdf}\end{eqnarray}
Estimates of $\overline{s}$, $u_{s}^{2}$ and $u_{mod}^{2}$ are
obtained from this pdf. In the general case, this has to be done numerically.\citep{Gregory05}
Two limit cases of interest (\emph{i.e.} negligible and very large
measurement uncertainties), amenable to analytical results, are presented
in the next sections.

\subsubsection{The case of negligible measurement uncertainties}

In the commonly met situation where the amplitude of the discrepancy
between calibration model and experimental data is much larger than
any other sources of uncertainty ($u_{mod}\gg u_{exp}$), we can consider
the approximate calibration model\begin{equation}
\nu_{exp,i}=s\omega_{i}+\epsilon_{mod},\label{eq:meas_eq}\end{equation}
for which the posterior pdf (Eq. \ref{eq:full_pdf}) can be simplified
and rearranged to (see Appendix \ref{sub:Case-of-negligible})\begin{eqnarray}
p\left(s,u_{mod}|\left\{ \nu_{exp,i},\omega_{i}\right\} _{i=1}^{N}\right) & \propto & \frac{1}{u_{mod}^{N+1}}\,\exp\left(-\frac{N\gamma^{2}}{2u_{mod}^{2}}\right)\,\exp\left(-\frac{(s-\hat{s})^{2}\sum_{i=1}^{N}\omega_{i}^{2}}{2u_{mod}^{2}}\right),\end{eqnarray}
from which one can analytically derive estimates of the parameters:

\begin{itemize}
\item $\overline{s}=\hat{s}$: the average value for $s$ is identical to
the optimal value of least-squares analysis (Eq. \ref{eq:optim_s});
\item $u_{s}$ , the standard uncertainty on $s$, is related to the rms
$\gamma$ by \begin{equation}
u_{s}=\gamma\,\sqrt{N/\left[(N-3)\sum\omega_{i}^{2}\right]};\label{eq:us}\end{equation}

\item and the estimate of $u_{mod}^{2}$ is related to $\gamma$ according
to \begin{equation}
\overline{u_{mod}^{2}}=\gamma^{2}\, N/(N-3).\label{eq:umod}\end{equation}

\end{itemize}
Inserting these values in Eq. \ref{eq:UP_glob}, we obtain the standard
uncertainty of a predicted frequency: \begin{equation}
u_{\nu}=\gamma\,\sqrt{\frac{N}{N-3}\left(\frac{\omega^{2}}{\sum_{i}\omega_{i}^{2}}+1\right)}.\label{eq:additive_UP}\end{equation}
It can be seen that for large calibration sets of few hundreds of
frequencies $\sqrt{N/(N-3)}\simeq1$ and $\omega^{2}/\sum_{i}\omega_{i}^{2}\ll1$,
and thus \begin{equation}
u_{\nu}\simeq\gamma.\label{eq:UP_simple}\end{equation}
In such conditions, it is possible to derive directly confidence intervals
on scaled properties from the summary calibration statistics $\hat{s}$
and $\gamma$ typically provided in the literature.\citep{Scott96,Wong96,Merrick07}
Assuming the normality of uncertainty distributions, confidence intervals
can be defined for prediction purpose, \emph{e.g. } the 95\% confidence
interval for $\nu$ is given by\begin{equation}
CI_{95}(\nu)=[\hat{s}\omega-1.96\, u_{\nu},\hat{s}\omega+1.96\, u_{\nu}].\label{eq:CI95}\end{equation}

\subsubsection{The case of very large measurement uncertainties\label{sub:Case-of-very}}

When model discrepancy is negligible compared to measurement uncertainties
($u_{mod}\ll u_{exp}$), the standard linear model is statistically
valid, and one recovers the Bayesian version of weighted least squares.
The posterior pdf for $s$ is then\begin{eqnarray}
p(s|\left\{ \nu_{exp,i},u_{exp,i},\omega_{i}\right\} _{i=1}^{N}) & \propto & \prod_{i=1}^{N}u_{exp,i}^{-1}\exp\left(-\frac{1}{2}\sum_{i=1}^{N}\frac{\left(\nu_{exp,i}-s\omega_{i}\right)^{2}}{u_{exp,i}^{2}}\right),\end{eqnarray}
from which one obtains\begin{align}
\hat{s} & =\sum_{i=1}^{N}\left(\omega_{i}\nu_{exp,i}/u_{exp,i}^{2}\right)/\sum_{i=1}^{N}\left(\omega_{i}^{2}/u_{exp,i}^{2}\right),\label{eq:wls}\\
u_{s}^{2} & =1/\sum_{i=1}^{N}\left(\omega_{i}^{2}/u_{exp,i}^{2}\right).\label{eq:us_wls}\end{align}
For uniform experimental uncertainty over the dataset, the scaling
factor uncertainty is \begin{equation}
u_{s}=u_{exp}/\sqrt{\sum_{i=1}^{N}\omega_{i}^{2}}.\label{eq:wls_lin}\end{equation}

\subsection{The Multiplicative Uncertainty (MU) method}

Irikura \emph{et al.} \citep{Irikura05}, after a thorough analysis
of the uncertainty sources in the \emph{ab initio} calculation of
harmonic vibrational frequencies, proposed that the major contribution
to prediction uncertainty would be the uncertainty on the scaling
factor $\hat{s}$. They estimate $u_{s}$ from the weighted variance
of $s$ with weights $a_{i}=\omega_{i}^{2}$. This weighting scheme
is derived in two steps: (1) they propose that the probability density
function (pdf) for the scaling factor is a linear combination of pdf's
for individual scaling factors in the reference set; and (2) from
the comparison of the expression of the average value derived from
this proposition with the least-squares solution Eq. \ref{eq:optim_s}.
This way, they obtain a standard uncertainty\begin{equation}
u_{s}^{*}\simeq\left(\frac{1}{\sum\omega_{i}^{2}}\sum\omega_{i}^{2}\left(s_{i}-\hat{s}\right)^{2}\right)^{1/2},\label{eq:us_irikura}\end{equation}
which can be related to the rms $\gamma$ by $u_{s}^{*}\simeq\gamma\,\sqrt{N/\sum\omega_{i}^{2}}$. 

More recently, Irikura \emph{et al.} \citep{Irikura09} derived another
expression by standard uncertainty propagation from the least-squares
solution Eq. \ref{eq:optim_s}, adding a new term to their previous
expression\begin{equation}
u_{s}^{*}\simeq\left(\frac{1}{\sum\omega_{i}^{2}}\sum\omega_{i}^{2}\left(s_{i}-\hat{s}\right)^{2}+\frac{1}{\left(\sum\omega_{i}^{2}\right)^{2}}\sum\omega_{i}^{2}u_{exp,i}^{2}\right)^{1/2}.\end{equation}
They showed that the contribution of the latter term is negligible,
validating the use of their former expression. Note that, unless all
frequencies $\omega_{i}$ are equal, this uncertainty $u_{s}^{*}$
is different from the dispersion of $s$ values within the calibration
set\begin{equation}
\delta_{s}=\left(\frac{1}{N}\sum\left(s_{i}-\hat{s}\right)^{2}\right)^{1/2}\label{eq:sigma_s}\end{equation}
and attributes larger weights to the high frequencies.

Using either of Irikura \emph{et al.}\citep{Irikura05,Irikura09}
expressions for $u_{s}^{*}$, uncertainty on a scaled frequency is
approximated by\begin{equation}
u_{\nu}\simeq\omega u_{s}^{*},\label{eq:UP_irikura}\end{equation}
hence the name of ''Multiplicative Uncertainty'' (MU) used hereafter. 

The salient feature of Eq. \ref{eq:UP_irikura} is that prediction
uncertainty is always proportional to the calculated harmonic frequency,
ignoring the additive term present in Eq. \ref{eq:UP_glob}. Simple
statistical validation test of the MU method have apparently not been
published and are performed in the next sections.

\section{Applications and discussion\label{sec:Applications-and-discussion}}

In the following, we validate the BMC approach and compare it to the
MU approach on representative test cases of vibrational frequencies
and zero point energies.

\subsection{Vibrational frequencies}

The reference dataset of 2737 frequencies for the HF/6-31G{*} combination
of theory/basis-set has been downloaded from the NIST/CCCBDB in July
2008.\citep{cccbdb} Correlation between experimental and harmonic
frequencies is plotted in Fig. \ref{fig:Correlation-plot}.

\subsubsection{Calibration\label{sub:Calibration}}

In absence of detailed information on the measurement uncertainties
for this dataset, and considering the typical high accuracy of spectroscopic
data, we assume that they are negligible and apply the corresponding
equations for the BMC model. Using Eqs. \ref{eq:us} and \ref{eq:umod},
we obtain $\hat{s}=0.89843\pm0.00046$, and $u_{mod}=45.35\pm0.61$~cm$^{-1}$
(Table \ref{tab:full_set}). The latter value is very close to the
rms value $\gamma=45.33$~cm$^{-1}$, which validates the use of
Eq. \ref{eq:UP_simple} for large calibration datasets.

\begin{figure}
\noindent \begin{centering}
\includegraphics[clip,width=12cm]{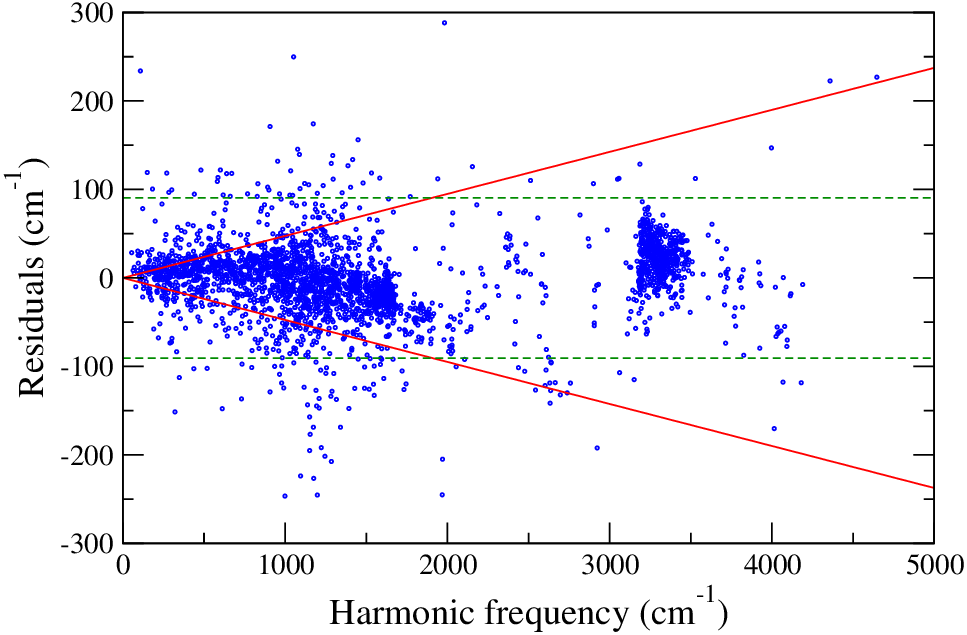}
\par\end{centering}

\caption{\label{fig:CI}Residuals of the linear scaling model for a set of
2737 vibrational frequencies and the HF/6-31G{*} combination of theory/basis-set
(dots). Model 95\,\% confidence intervals for residuals: dashed (green)
lines for the Bayesian Model Calibration method; solid (red) lines
for the Multiplicative Uncertainty model of IJK05/IJKK09.}

\end{figure}

For this same dataset, the CCCBDB proposes $\hat{s}=0.899\pm0.025$,
which can be recovered using Eq. \ref{eq:us_irikura} (Table \ref{tab:full_set}).
The standard uncertainties on $\hat{s}$ evaluated by both methods
differ thus by a factor 50, which can be expected to have noticeable
effect on prediction uncertainty (see Section \ref{sub:Uncertainty-propagation}).
In order to visualize the difference, we plotted the 95\,\% confidence
intervals on predicted frequencies obtained from both methods (Fig.
\ref{fig:CI}). It is immediately visible that the the MU approach
has a tendency to underestimate uncertainty at low frequencies and
to overestimate it at high frequencies. 

\begin{table}
\noindent \begin{centering}
\begin{tabular}{lccccccccc}
\toprule 
 & \multicolumn{2}{c}{Summary stat.} &  & \multicolumn{2}{c}{MU} &  & \multicolumn{3}{c}{BMC}\tabularnewline
\cmidrule{2-3} \cmidrule{5-6} \cmidrule{8-10} 
 & $\hat{s}$ & $\gamma$(cm$^{-1}$) &  & $u_{s}^{*}$ & \%CI$_{95}$ &  & $u_{s}$ & $u_{mod}$(cm$^{-1}$) & \%CI$_{95}$\tabularnewline
\midrule
\multicolumn{10}{l}{\textbf{All frequencies} ($N=2737$)}\tabularnewline
Full set  & 0.89843 & 45.33 &  & 0.025 & - &  & 0.00046 & 45.35 & -\tabularnewline
Calibration set & 0.89860 & 45.27 &  & 0.024 & - &  & 0.00065 & 45.31 & -\tabularnewline
Validation set & - & - &  & - & 83.0 &  & - & - & 94.6\tabularnewline
\midrule
\multicolumn{10}{l}{\textbf{High frequencies, between 3180 and 3500 cm$^{-1}$} ($N=479$)}\tabularnewline
Full set  & 0.90502 & 28.71 &  & 0.00869 & - &  & 0.00040 & 28.78 & -\tabularnewline
Calibration set & 0.90517 & 23.32 &  & 0.01005 & - &  & 0.00046 & 23.44 & -\tabularnewline
Validation set & - & - &  & - & 97.4 &  & - & - & 95.4\tabularnewline
\bottomrule
\end{tabular}\caption{\label{tab:full_set}Statistical estimates and validation for MU and
BMC models for vibrational frequencies extracted from the CCCBDB for
the HF/6-31G{*} combination of theory/basis-set.}

\par\end{centering}
\end{table}

\subsubsection{Validation}

To better quantify this inconsistency, we performed a standard test
in statistical calibration/prediction: the dataset is split randomly
in two subsets, one for calibration, the other one for validation.
Both sets are taken here of equal size (plus or minus one unit). In
this case, one gets slightly different values of the parameters, as
reported in Table \ref{tab:full_set}. Using these values, we generate
95\,\% confidence intervals (Eq. \ref{eq:CI95}; the residuals of
this dataset have a nearly normal distribution) and calculate the
percentage of inclusion of the experimental values of the validation
subset within these prediction intervals (Fig. \ref{fig:CI}). For
a consistent predictor, one should find a frequency close to 95\,\%.
BMC succeeds for 94.7\,\% of the frequencies in the validation set,
whereas the MU model succeeds for only 83\,\% (Table \ref{tab:full_set}).
Considering the size of the samples, the difference is significant,
and the statistical consistency of the MU approach can be questioned.
When contrasted with the BMC, one understands that the MU method,
which does not consider model inadequacy explicitly, ''absorbs''
it at least partially in $u_{s}^{*}$.

\subsubsection{Test on a restricted frequency scale}

As stated in IJK05, ''to apply the fractional bias correction, it
is important to select a class of frequencies similar to the ones
to be estimated''.\citep{Irikura05} For instance, if one selects
in the reference set only those frequencies between 3180 and 3500
cm$^{-1}$, one gets a much more uniform picture than with the full
reference set. 

The MU calibration procedure was done with this limited set of 479
frequencies, providing $\hat{s}=0.9050\pm0.0087$ (Table \ref{tab:full_set}).
In this case, the uncertainty factor for $s$ is practically identical
to the standard deviation calculated from the sample (0.00869 vs.
0.00871): $u_{s}^{*}\simeq\delta_{s}$. Due to the restricted frequency
range, one has indeed  $\omega_{i}^{2}/\sum_{i}\omega_{i}^{2}\simeq1/N$
\emph{, }hence the identity between evaluations by Eqs \ref{eq:us_irikura}
and \ref{eq:sigma_s}. 

This set has been split in two, as before. The scaling factor obtained
by MU from the calibration subset is now $\hat{s}=0.9052\pm0.0071$,
and 97.4\,\% of the validation frequencies fall within the 95\,\%
confidence interval. This result is quite close to the one obtained
with BMC (Table \ref{tab:full_set}).

It appears thus that in restrictive conditions, the MU method can
be valid for reference sets where the individual scaling factors are
uniformly distributed with regard to the harmonic frequencies. In
such cases however, the uncertainty is recovered as the conventional
unweighted standard deviation of the sample of individual scaling
factors (Eq. \ref{eq:sigma_s}). Note also that the MU method is used
in the CCCBDB out of these favorable conditions.\citep{cccbdb}

\subsubsection{Significant figures and uncertainty reporting}

Good practice in uncertainty reporting is to provide one or two significant
figures for the uncertainty and to truncate the average/optimal value
at the same level.\citep{GUM} If the reported number is to be used
in further calculations (which is the case for uncertainty propagation),
two digits is better. The common practice is to publish scaling factors
for vibrational frequencies with four significant digits.\citep{Scott96,Wong96,Merrick07}
At the risk of being pedantic, one could argue that they should be
reported with \emph{five} significant digits, e.g. $\hat{s}=0.89843\pm0.00046$,
in sharp contrast with the two digits recommendation of Irikura \emph{et
al.},\citep{Irikura05} based on their biased scaling factor uncertainty
evaluation.

\subsubsection{Prediction and uncertainty propagation\label{sub:Uncertainty-propagation}}

The relative importance of both factors $u_{s}^{2}$ and $u_{mod}^{2}$
in Eq. \ref{eq:UP_glob} can be evaluated on the example of a calculated
harmonic frequency in the higher frequency range $\omega=3000$ cm$^{-1}$
(Table \ref{tab:UP}). 

In this case, the uncertainty on the scaling factor contributes only
to one thousandth of the total prediction variance. When dealing with
large datasets of accurate vibrational frequencies, the uncertainty
on the scaling factor can thus be neglected. The uncertainty on $u_{mod}$
is also much too small to be relevant for confidence intervals calculation.
One can therefore safely apply the uncertainty propagation formula
(Eq. \ref{eq:UP_simple}), using the rms provided by most reference
articles dealing with scaling factors calibration.\citep{Scott96,Wong96,Merrick07}
For smaller calibration sets, the rms can be seen as an inferior limit
to prediction uncertainty, and Eq. \ref{eq:additive_UP} would provide
more reliable confidence intervals (see next Section).

Comparing the prediction uncertainties for the BMC (45\,cm$^{-1}$)
and MU (75\,cm$^{-1}$) methods, one sees that the factor 50 between
$u_{s}$ and $u_{s}^{*}$ observed at the calibration stage is partially
damped at the prediction level by the fact that the BMC uncertainty
is strongly dominated by the model inadequacy parameter $u_{mod}$. 

\begin{center}
\begin{table}
\noindent \begin{centering}
\begin{tabular}{lllllll}
\hline 
Property & Theory/Basis set & $\omega$  & Method & $\omega^{2}u_{s}^{2}$ & $u_{mod}^{2}$ & $\nu\pm u_{\nu}$\tabularnewline
\hline
Frequency  & HF/6-31G{*} & 3000~cm$^{-1}$ & BMC & 2.25 & 2052.09 & 2695$\pm$45~cm$^{-1}$\tabularnewline
 &  &  & MU & 5625.0 & - & 2695$\pm$75~cm$^{-1}$\tabularnewline
\hline
ZPE  & HF/6-31G{*} & 100~kJ~mol$^{-1}$ & BMC & 0.073 & 0.53 & 91.35$\pm$0.78~kJ~mol$^{-1}$\tabularnewline
 &  &  & MU & 2.592 & - & 91.35$\pm$1.61~kJ~mol$^{-1}$\tabularnewline
\hline
ZPE  & B3LYP/6-31G{*} & 100~kJ~mol$^{-1}$ & BMC & 0.029 & 0.19 & 98.12$\pm$0.47~kJ~mol$^{-1}$\tabularnewline
 &  &  & MU & 1.061 & - & 98.12$\pm$1.03~kJ~mol$^{-1}$\tabularnewline
\hline
\end{tabular}\caption{\label{tab:UP}Compared prediction uncertainty with the BMC and MU
methods for a set of 2737 vibrational frequencies extracted from the
CCCBDB for the HF/6-31G{*} combination of theory/basis-set and for
a set of 39 ZPE of the Z1 set for the HF/6-31G{*} and B3LYP/6-31G{*}
combinations.}

\par\end{centering}
\end{table}

\par\end{center}

\subsection{Zero Point Vibrational Energies}

We consider ZPE as an additional test because the reference datasets
are considerably smaller than for the vibrational frequencies (e.g.
39 molecules in the Z1 set of Merrick \emph{et al.} \citep{Merrick07}),
which is expected to emphasize the role of $u_{s}$, the uncertainty
on the scaling factor. The uncertainties reported by Irikura \citep{Irikura07}
for diatomic molecules are typically very small (on the order of 0.01
cm$^{-1}$), but transposition to larger molecules is not straightforward.
In the absence of a systematic review of measurement errors for ZPE
of polyatomic molecules, we consider here that they can be neglected.
The effect of non-negligible measurement uncertainties is addressed
at the end of this section.

\subsubsection{Calibration - Validation}

Using BMC with the Z1 reference set, one gets $\hat{s}=0.9135\pm0.0027$
and $u_{mod}=0.731\pm0.086$~kJ~mol$^{-1}$ (Table \ref{tab:Z1}),
which is consistent with the rms obtained by Merrick \emph{et al.}
\citep{Merrick07} for the HF/6-31G{*} theory/basis-set combination.
Relative uncertainties on these parameters have been increased by
one order of magnitude, when compared to the vibrational frequencies
case, a direct effect of the smaller sample size. The validation test
shows once more that the MU model fails to provide correct confidence
intervals, with a score of only 0.63 for CI$_{95}$ (Table \ref{tab:Z1}).

\subsubsection{Uncertainty propagation}

For such a small reference dataset, it is interesting to check if
the approximate formula (Eq. \ref{eq:UP_simple}) for uncertainty
propagation, which was validated for large sets of vibrational frequencies
still holds, \emph{i.e.} if the contribution of the multiplicative
term involving $u_{s}$ stays negligible or not for the larger ZPE
values. If one considers a calculated ZPE of 100~kJ~mol$^{-1}$
(HF/6-31G{*}), one has $u_{\nu}=\sqrt{(100*0.0027)^{2}+0.73^{2}}=0.78$~kJ~mol$^{-1}$,
to be compared to $\gamma=0.71$~kJ~mol$^{-1}$(Table \ref{tab:Z1}).
It is to be noted also that the uncertainty on $u_{mod}$ might also
contribute, with $u_{u_{mod}}=0.09$ kJ~mol$^{-1}$. Taking all uncertainty
sources into account trough Eq. \ref{eq:UP_exact} by Monte Carlo
Uncertainty Propagation (MCUP),\citep{GUMSupp1} one gets $u_{\nu}=0.77$~kJ~mol$^{-1}$.
The uncertainty on $u_{mod}$ can therefore be neglected.

In the same conditions, for the combination B3LYP/6-31G{*}, one gets
$\gamma=0.42$~kJ~mol$^{-1}$ and $u_{\nu}=0.47$~kJ~mol$^{-1}$,
to be compared with a reference value obtained by MCUP of $u_{\nu}=0.47$~kJ~mol$^{-1}$
(Table \ref{tab:Z1}).

There is globally only a 10\% increase compared to the rms $\gamma$.
In this range of ZPEs, $\gamma$ still provides a good approximation
of the prediction uncertainty (Table \ref{tab:UP}). However, the
amplitude of the discrepancy between $\gamma$ and $u_{\nu}$ will
probably increase with the size of the molecule. In consequence, for
uncertainty propagation with ZPEs, notably for large molecules, it
would be safer to use the full UP formula (Eq. \ref{eq:additive_UP}),
involving the multiplicative uncertainty factor. Compilations of scaling
factors for ZPE should thus report the easily calculated value of
$u_{s}=\gamma\,\sqrt{N/\left((N-3)\sum\omega_{i}^{2}\right)}$, in
addition to the rms $\gamma$.

\begin{table}
\noindent \begin{centering}
\begin{tabular}{lccccccccc}
\toprule 
 & \multicolumn{2}{c}{Summary stat.} &  & \multicolumn{2}{c}{MU} &  & \multicolumn{3}{c}{BMC}\tabularnewline
\cmidrule{2-3} \cmidrule{5-6} \cmidrule{8-10} 
 & $\hat{s}$ & $\gamma$(kJ~mol$^{-1}$) &  & $u_{s}^{*}$ & \%CI$_{95}$ &  & $u_{s}$ & $u_{mod}$(kJ~mol$^{-1}$) & \%CI$_{95}$\tabularnewline
\midrule
\multicolumn{10}{l}{\textbf{HF/6-31G{*} }}\tabularnewline
Full set  & 0.9135 & 0.707 &  & 0.0161 & - &  & 0.0027 & 0.731$\pm$0.086 & -\tabularnewline
Calibration set & 0.9078 & 0.773 &  & 0.0214 &  &  & 0.0052 & 0.826$\pm$0.143 & \tabularnewline
Validation set &  &  &  &  & 0.63 &  &  &  & 0.95\tabularnewline
\midrule
\multicolumn{10}{l}{\textbf{B3LYP/6-31G{*}}}\tabularnewline
Full set  & 0.9812 & 0.423 &  & 0.0103 & - &  & 0.0017 & 0.437$\pm$0.052 & -\tabularnewline
Calibration set & 0.9825 & 0.448 &  & 0.0134 &  &  & 0.0032 & 0.478$\pm$0.083 & \tabularnewline
Validation set &  &  &  &  & 0.63 &  &  &  & 1.00\tabularnewline
\bottomrule
\end{tabular}\caption{\label{tab:Z1}Statistical estimates and validation for MU and BMC
models for a set of 39 ZPVEs of the Z1 set for the HF/6-31G{*} and
B2LYP/6-31G{*} combinations of theory/basis-set.}

\par\end{centering}
\end{table}

\subsubsection{The case of non-negligible experimental uncertainties}

\begin{figure}
\noindent \begin{centering}
\includegraphics[clip,width=12cm]{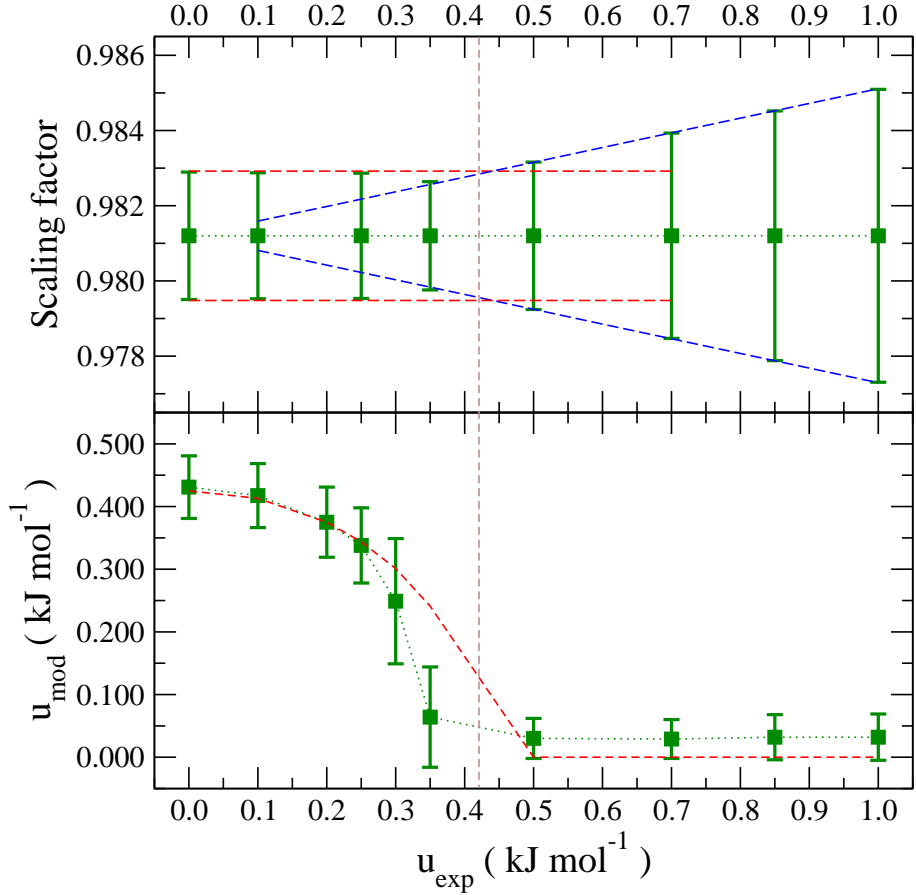}
\par\end{centering}

\caption{\label{fig:Uncert_sc}Evolution of measurement model parameters with
the amplitude of an hypothetical uniform experimental measurement
uncertainty $u_{exp}$ on ZPE; B3LYP/6-31G{*} combination of theory/basis-set
(green squares with error bars). The brown vertical dashed line indicates
the value of the rms $\gamma$. Top panel: the red dashed lines represent
the 1$\sigma$ confidence interval in the limit of null experimental
uncertainty; the blue dashed line represent the 1$\sigma$ confidence
interval in the weighted least squares limit. Bottom panel: the red
dashed line represents the $u_{mod}^{2}+u_{exp}^{2}=\gamma^{2}$ law,
truncated to positive values of $u_{mod}$.}

\end{figure}

When the measurement uncertainty becomes comparable to the rms, model
inadequacy should be small, and confidence intervals for prediction
should account for the measurement uncertainty (Eq. \ref{eq:UP_glob}).
In the absence of an exhaustive compilation of experimental uncertainties
on measured ZPE, we performed simulations assuming a uniform uncertainty
distribution over the full dataset. In order to test the sensitivity
of the model parameters to this uncertainty, we repeated the estimations
of the previous section, using Eq. \ref{eq:bayes_exp_unc}, for values
of $u_{exp}$ between 0.1 and 1.0~kJ~mol$^{-1}$. The results are
reported in Fig. \ref{fig:Uncert_sc}.

As expected from the properties of the posterior pdf (Eq. \ref{eq:full_pdf}),
the average/optimal value of the scaling factor is not sensitive to
the amplitude of $u_{exp}$. Moreover, we observe only a slight absolute
increase of $u_{s}$ from 0.002 to 0.004. A transition from a constant
$u_{s}$, defined by the $u_{exp}=0$ limit, to a linear increase
consistent with the weighted least squares limit (Eq. \ref{eq:wls_lin})
is observed around $u_{exp}=\gamma$, where both limit equations intersect.
A closer look shows that the transition occurs indeed at values of
$u_{exp}$ slightly smaller than $\gamma$, in a region ($u_{exp}\simeq0.35)$
where $u_{s}$ displays a minimum. 

The evolution of the model inadequacy factor $u_{mod}$ is more dramatic:
it displays a sharp decrease and falls down to zero as soon as the
measurement uncertainty reaches the value of the rms $\gamma$. For
values of $u_{exp}$ below $0.25$, $u_{mod}$ follows the $u_{mod}^{2}+u_{exp}^{2}=\gamma^{2}$
law (represented as a dashed line in Fig. \ref{fig:Uncert_sc}), but
the calculated decrease becomes much faster in the transition zone.
The uncertainty on $u_{mod}$ increases notably in the transition
region.

In the limit of large experimental uncertainties, using Eq. \ref{eq:wls_lin},
the uncertainty propagation formula can be written as \begin{align}
u_{\nu}^{2} & =\omega^{2}u_{s}^{2}=u_{exp}^{2}\omega^{2}/\sum\omega_{i}^{2}.\end{align}
In this case, the model inadequacy variable $\epsilon_{mod}$ becomes
useless, as the standard calibration model is statistically valid.

This test shows that the BMC model is able to adapt nicely to various
conditions of measurement uncertainty, with an automatic and smooth
transition from the ''model inadequacy''- to the ''measurement
uncertainty''-dominated regimes.

\section{Conclusions and recommendations}

A reanalysis of the scaling factor calibration problem as stated by
Irikura \emph{et al.} \citep{Irikura05,Irikura09} identified two
uncertainty components, besides the experimental one: a parametric
uncertainty $u_{s}$ attached to the optimal scaling factor, and a
model inadequacy factor $u_{mod}$ accounting for the inability of
the linear scaling correction model to reproduce sets of calibration
data within their experimental uncertainties. A general estimation
framework, based on Bayesian Model Calibration, has been defined and
validated in cases of interest. 

The general formula for prediction of a scaled property $\nu$ from
a harmonic value $\omega$ is\begin{equation}
\nu=\hat{s}\omega\pm u_{\nu},\end{equation}
where $\hat{s}$ is the optimal value of the scaling factor provided
by the least squares formula (Eq. \ref{eq:optim_s}) for negligible
or uniform measurement uncertainty, or more generally by the weighted
least squares formula (Eq. \ref{eq:wls}), and $u_{\nu}$ is a standard
uncertainty, for which explicit expressions have been derived in limit
cases, depending on the size and precison of the calibration set:

\begin{itemize}
\item large calibration sets of precise data ($u_{exp}\ll\gamma$): $u_{\nu}(\omega)=\gamma$;
\item small calibration sets of precise data ($u_{exp}\ll\gamma$): $u_{\nu}(\omega)=\gamma\,\sqrt{\frac{N}{N-3}\left(1+\omega^{2}/\sum_{i}\omega_{i}^{2}\right)}$;
\item calibration sets with large measurement uncertainties ($u_{exp}\geq\gamma$):
$u_{\nu}(\omega)=\omega/\sqrt{\sum_{i}\omega_{i}^{2}/u_{exp,i}^{2}}$,
simplified to $u_{\nu}(\omega)=u_{exp}\omega/\sqrt{\sum_{i}\omega_{i}^{2}}$
for uniform measurement uncertainty.
\end{itemize}
The Multiplicative Uncertainty method proposed by Irikura \emph{et
al.}\citep{Irikura05,Irikura09} has been shown here to be statistically
inconsistent when large frequency ranges are considered. It is only
valid in particular situations, either when the dataset spans a restricted
frequency range (in which case the uncertainty is reduced to a trivial
unweighted standard deviation), or in the extreme case of large uniform
measurement uncertainty in the calibration dataset. For vibrational
frequencies, the MU method underestimates prediction uncertainty for
small values of $\omega$ and overestimate it (up to a factor 2) at
the high end of the $\omega$ scale. 

We would like to stress out that the validity of the formulas proposed
above for uncertainty propagation depends to some extent on the normality
of the residuals $\left\{ \nu_{i}-\hat{s}\omega_{i}\right\} _{i=1}^{N}$
of the linear regression. Inspection of histograms of residuals (see
\emph{e.g.} Fig. 1 in Ref. \citep{Wong96}) shows that this is not
always the case. The usual approach of choosing an optimal theory/basis-set
combination is to assess their performance by the rms alone, maybe
weighted by computational cost considerations.\citep{Scott96,Wong96,Merrick07}
Researchers concerned by prediction uncertainty might also consider
an additional ''normality criterion'' in order to reject theory/basis-set
combinations providing non-normal residuals and from which the summary
statistics cannot be used reliably for uncertainty propagation. Analysis
of restricted ranges of data as presently done by some authors for
vibrational frequencies\citep{Merrick07,Bouteiller09} is one way
to improve the normality of residuals, but as demonstrated above,
prediction from small calibration sets calls for more information
than the rms.

\subsection{Recommendations to calibrators of scaling factors}

\begin{enumerate}
\item For large calibration sets of accurate data, as the ones used for
calibration of uniform scaling factors for vibrational frequencies,
reliable prediction uncertainty can be simply based on the rms $\gamma$
(Eq. \ref{eq:rms}). In this case, prediction uncertainty is purely
additive.
\item For much smaller datasets of a few dozens of data or less, as in the
case of ZPEs or mode-specific frequencies, one has a combination of
additive and multiplicative uncertainty (or rather, variance). Ideally,
uncertainty on the scaling factor $u_{s}=\gamma\,\sqrt{N/\left((N-3)\sum\omega_{i}^{2}\right)}$
should be reported along with the additive term $u_{mod}=\gamma\,\sqrt{N/(N-3)}$,
for use in the general uncertainty propagation equation $u_{\nu}^{2}=\omega^{2}u_{s}^{2}+u_{mod}^{2}$
(Eq. \ref{eq:UP_glob}). It certainly would be a large step towards
the general applicability of the Virtual Measurement concept,\citep{Irikura04}
if statistically pertinent estimators were systematically reported
in the literature devoted to the calibration of semi-empirical correction
parameters.
\item An indicator of the normality of the residuals in the calibration
dataset would also be welcomed.
\end{enumerate}

\subsection{Recommendations to users of scaling factors}

\begin{enumerate}
\item For the end user of scaling factors, it is important to remind, as
pertinently stated by Irikura \emph{et al.}\citep{Irikura05,Irikura09},
that semi-empirical correction of a property by scaling is not a deterministic
procedure: a scaled property has an attached uncertainty, which depends
on the level of theory/basis-set used for the calculation of harmonic
properties (it depends also on the quantity and quality of the calibration
dataset, but this is out of reach of the end user). 
\item In the present state of affairs, the best estimate of the prediction
uncertainty available for most levels of theory/basis-set is the rms
$\gamma$, \emph{i.e. }one has to assume $u_{\nu}\simeq\gamma$.\citep{Scott96,Wong96,Merrick07}
The use of the multiplicative scaling factor uncertainty as reported
presently (March 2010) in the CCCBDB\citep{cccbdb} cannot be recommended
for the estimation of uncertainty of scaled properties.
\item Users are encouraged to 

\begin{enumerate}
\item report the uncertainty along with the scaled properties, \emph{i.e.}
$\nu=\hat{s}\omega\pm u_{\nu}$, and 
\item account for uncertainty when scaled properties are used as inputs
to a model\citep{GUM,GUMSupp1}, or for comparison with experimental
data.
\end{enumerate}
\item One has to be conscious that $\gamma$ provides only a lower limit
of the uncertainty for properties with small calibration data sets
(\emph{e.g.} ZPE). For numerical examples, see Table \ref{tab:UP}.
\end{enumerate}

\section*{Acknowledgments}

The authors would like to thank Prof. Leo Radom for providing the
Z1 ZPE dataset. B. L\'evy is warmly acknowledged for helpful discussions.

\appendix

\section{Appendix}

\subsection{Bayesian analysis of scaling factor calibration model\label{sub:Taking-measurement-uncertainties}}

We consider the calibration model \begin{equation}
\nu_{exp,i}=s\omega_{i}+\epsilon_{mod}+\epsilon_{exp,i},\end{equation}
where $\epsilon_{exp,i}\sim N(0,u_{exp,i}^{2})$ is the measurement
uncertainty of $\nu_{exp,i}$, and $\epsilon_{mod}\sim N(0,u_{mod}^{2})$
is a variable accounting for the discrepancy between the linear model
and the observations. This model has two unknown parameters, $s$
and $u_{mod}$, to be estimated on a calibration dataset consisting
of $N$ calculated harmonic frequencies $\left\{ \omega_{i}\right\} _{i=1}^{N}$,
and their corresponding experimental frequencies $\left\{ \nu_{exp,i},u_{exp,i}\right\} _{i=1}^{N}$.

In the Bayesian data analysis framework,\citep{Sivia96,Gregory05}
all information about parameters can be obtained from the joint posterior
pdf $p\left(s,u_{mod}|\left\{ \nu_{exp,i},u_{exp,i},\omega_{i}\right\} _{i=1}^{N}\right)$.
In order to simplify the notations, we will omit in the following
the list indices when they are not necessary. This pdf is obtained
through Bayes theorem\begin{eqnarray}
p(s,u_{mod}|\left\{ \nu_{exp},u_{exp},\omega\right\} ) & \propto & p(\left\{ \nu_{exp}\right\} |s,u_{mod},\left\{ u_{exp},\omega\right\} )\, p(s,u_{mod}),\label{eq:Bayes}\end{eqnarray}
where $p(\left\{ \nu_{exp}\right\} |s,u_{mod},\left\{ u_{exp},\omega\right\} )$
is the likelihood function and $p(s,u_{mod})$ is the prior pdf. 

In the hypothesis where the difference between observation and corrected
frequency is expected to arise from a normal distribution \begin{equation}
\nu_{exp,i}-s\omega_{i}\sim N(0,u_{mod}^{2}+u_{exp,i}^{2}),\end{equation}
the likelihood function for a single observed frequency is\begin{equation}
p(\nu_{exp,i}|s,u_{mod},u_{exp,i},\omega_{i})=\left(2\pi\left(u_{mod}^{2}+u_{exp,i}^{2}\right)\right)^{-1/2}\exp\left(-\frac{1}{2}\frac{\left(\nu_{exp,i}-s\omega_{i}\right)^{2}}{u_{mod}^{2}+u_{exp,i}^{2}}\right).\end{equation}
Considering that all frequencies are measured independently (with
uncorrelated uncertainty) the joint likelihood is the product of the
individual ones, \emph{i.e.}\begin{eqnarray}
p\left(\left\{ \nu_{exp}\right\} |s,u_{mod},\left\{ u_{exp},\omega\right\} \right) & = & \prod_{i=1}^{N}\left(2\pi\left(u_{mod}^{2}+u_{exp,i}^{2}\right)\right)^{-1/2}\times\nonumber \\
 &  & \exp\left(-\frac{1}{2}\sum_{i=1}^{N}\frac{\left(\nu_{exp,i}-s\omega_{i}\right)^{2}}{u_{mod}^{2}+u_{exp,i}^{2}}\right).\end{eqnarray}
As there is a priori no correlation between $s$ and $u_{mod}$, we
use a factorized prior pdf $p(s,u_{mod})=p(s)p(u_{mod})$. In the
absence of a priori quantified information on $s$, a uniform pdf
$p(s)=cte$ is used. For $u_{mod}$, we enforce a positivity constraint
through a Jeffrey's prior, $p(u_{mod})\propto u_{mod}^{-1}$.\citep{Gregory05}
The posterior pdf is finally defined up to a norm factor which is
irrelevant for the following developments\begin{eqnarray}
p(s,u_{mod}|\left\{ \nu_{exp},\omega,u_{exp}\right\} ) & \propto & u_{mod}^{-1}\prod_{i=1}^{N}\left(u_{mod}^{2}+u_{exp,i}^{2}\right)^{-1/2}\times\nonumber \\
 &  & \exp\left(-\frac{1}{2}\sum_{i=1}^{N}\frac{\left(\nu_{exp,i}-s\omega_{i}\right)^{2}}{u_{mod}^{2}+u_{exp,i}^{2}}\right).\label{eq:bayes_exp_unc}\end{eqnarray}

\subsection{Case of negligible measurement uncertainties\label{sub:Case-of-negligible}}

For the analysis of vibrational frequencies, it is generally considered
that experimental uncertainties are negligible when compared to model
inadequacy ($u_{exp,i}\ll u_{mod}$). The general expression for the
posterior pdf (Eq. \ref{eq:bayes_exp_unc}) can then be simplified
accordingly:\begin{eqnarray}
p(s,u_{mod}|\left\{ \nu_{exp},\omega\right\} ) & \propto & u_{mod}^{-N-1}\,\exp\left(-\frac{1}{2u_{mod}^{2}}\sum_{i=1}^{N}\left(\nu_{exp,i}-s\omega_{i}\right)^{2}\right).\end{eqnarray}
Using Eq. \ref{eq:optim_s} and \ref{eq:rms} we derive the identity
(see \emph{e.g. }Ref. \citep[, Eq. 9.4, p. 214]{Gregory05}) \begin{equation}
\sum_{i=1}^{N}\left(\nu_{exp,i}-s\omega_{i}\right)^{2}=(s-\hat{s})^{2}\sum\omega_{i}^{2}+N\gamma^{2},\label{eq:identity}\end{equation}
which enables to write the posterior pdf in a convenient factorized
form\begin{eqnarray}
p(s,u_{mod}|\left\{ \nu_{exp},\omega\right\} ) & \propto & u_{mod}^{-N-1}\,\exp\left(-\frac{N\gamma^{2}}{2u_{mod}^{2}}\right)\,\exp\left(-\frac{(s-\hat{s})^{2}\sum\omega_{i}^{2}}{2u_{mod}^{2}}\right)\end{eqnarray}
from which we can derive analytical estimates for the parameters and
their uncertainties.

\subsubsection{Estimation of $s$}

The marginal density for $s$ is obtained by integration over $u_{mod}$

\begin{eqnarray}
p(s|\left\{ \nu_{exp},\omega\right\} ) & = & \int_{0}^{\infty}du_{mod}\, p(s,u_{mod}|\left\{ \nu_{exp},\omega\right\} )\\
 & \propto & \int_{0}^{\infty}du_{mod}\, u_{mod}^{-N-1}\exp\left(-\frac{1}{2u_{mod}^{2}}\sum_{i=1}^{N}\left(\nu_{exp,i}-s\omega_{i}\right)^{2}\right)\\
 & \propto & \left(\sum_{i=1}^{N}\left(\nu_{exp,i}-s\omega_{i}\right)^{2}\right)^{-N/2},\label{eq:like_1}\end{eqnarray}
which, using Eq. \ref{eq:identity}, can be rewritten as \begin{equation}
p(s|\left\{ \nu_{exp},\omega\right\} )\propto\left(1+\frac{(s-\hat{s})^{2}\sum\omega_{i}^{2}}{N\gamma^{2}}\right)^{-N/2},\end{equation}
and has the shape of a Student's distribution \citep{Evans00}

\begin{equation}
\mathrm{Stt}(x)\propto\left(1+\frac{x^{2}}{n}\right)^{-(n+1)/2}.\end{equation}
Posing $n=N-1$ and $x^{2}=(N-1)/N\,(s-\hat{s})^{2}\sum\omega_{i}^{2}/\gamma^{2}$,
we can directly use the properties of the Student's distribution \begin{equation}
\mathrm{E}[x]=0;\,\mathrm{Var}[x]=n/(n-2),\end{equation}
to derive\begin{align}
\mathrm{E}[s]\equiv\overline{s} & =\hat{s}\\
\mathrm{Var}[s]\equiv u_{s}^{2} & =\gamma^{2}\,\frac{N}{(N-1)\sum\omega_{i}^{2}}\mathrm{Var}[x]\\
 & =\gamma^{2}\,\frac{N}{(N-3)\sum\omega_{i}^{2}}\end{align}

\subsubsection{Estimation of $u_{mod}$}

The marginal density for the standard uncertainty of the stochastic
variable $\epsilon_{mod}$ is \begin{eqnarray}
p(u_{mod}|\left\{ \nu_{exp},\omega\right\} ) & = & \int_{-\infty}^{\infty}ds\, p(s,u_{mod}|\left\{ \nu_{exp},\omega\right\} )\\
 & \propto & \frac{1}{u_{mod}^{N+1}}\exp\left(-\frac{N\gamma^{2}}{2u_{mod}^{2}}\right)\int_{-\infty}^{\infty}ds\exp\left(-\frac{(s-\hat{s})^{2}\sum\omega_{i}^{2}}{2u_{mod}^{2}}\right)\\
 & \propto & \frac{1}{u_{mod}^{N}}\exp\left(-\frac{N\gamma^{2}}{2u_{mod}^{2}}\right).\end{eqnarray}
Using the formula\begin{equation}
\int_{0}^{\infty}dx\, x^{-n}e^{-a/x^{2}}=\frac{1}{2}\Gamma\left(\frac{n-1}{2}\right)/a^{(n-1)/2}\end{equation}
to recover the normalization constant of $p(u_{mod}|\left\{ \nu_{exp},\omega\right\} )$
and to calculate mean values of $u_{mod}$ and $u_{mod}^{2}$, one
obtains readily the following estimates\begin{align}
\hat{u}_{mod} & =\gamma\\
\overline{u}_{mod} & =\gamma\,\sqrt{\frac{N}{2}}\frac{\Gamma\left[(N-2)/2\right]}{\Gamma\left[(N-1)/2\right]}\\
\overline{u_{mod}^{2}} & =\frac{N}{N-3}\gamma^{2}\\
u_{u_{mod}} & =\gamma\,\sqrt{\frac{N}{N-3}-\frac{N}{2}\left(\frac{\Gamma\left[(N-2)/2\right]}{\Gamma\left[(N-1)/2\right]}\right)^{2}}.\end{align}

\subsection{Prediction and uncertainty propagation}

In the Bayesian framework, the posterior pdf $p(s,u_{mod}|\left\{ \nu_{exp},u_{exp},\omega\right\} )$
can be used to estimate the uncertainty of predicted frequencies by
the law of propagation of distribution\citep{GUMSupp1}\begin{equation}
p\left(\nu'|\omega',\left\{ \nu_{exp},u_{exp},\omega\right\} \right)=\int ds\, du_{mod}\, p(\nu'|s,u_{mod},\omega')p\left(s,u_{mod}|\left\{ \nu_{exp},u_{exp},\omega\right\} \right),\label{eq:UP_exact}\end{equation}
where\begin{equation}
p(\nu'|s,u_{mod},\omega')\propto u_{mod}^{-1}\exp\left(-\frac{\left(\nu'-s\omega'\right)^{2}}{2u_{mod}^{2}}\right)\end{equation}
translates the stochastic prediction model (Eq. \ref{eq:stoch_pred})
as a pdf . This integral has generally to be evaluated numerically.

\end{document}